\def\dvm {R^{1/m}}
\def\Iz {I_{\circ}}
\def\re {R_{\rm e}}
\def\ra {r_{\rm a}}
\def\sa {s_{\rm a}}
\def\sac{(\sa)_{\rm c}}
\def\sax{(\sa)_{\xi}}
\def\sp {\sigma_{\rm P}}
\def\srad{\sigma_{\rm r}}
\def\stan{\sigma_{\rm t}}
\def\sap{\sigma_{\rm a}}

\def\svir{\sigma_{\rm V}}
\def\Krad{K_{\rm r}}
\def\Ktan{K_{\rm t}}
\def\e {{\cal E}} 
\def\Qt{\widetilde Q}
\def\hq{h_4}

\def\ltot {L_m}
\def\ltota {\widetilde\ltot}
\def\pot {\psi _m}
\def\pota {\widetilde\pot}
\def\ml {\Upsilon}
\def\dm {\widetilde\nu _m}
\def\dfm{f_m}
\def\dfmt{\widetilde\dfm}
\def\dfqt{\widetilde {f_4}}
\catcode`\@=11
\def\gsim{\ifmmode{\mathrel{\mathpalette\@versim>}}
    \else{$\mathrel{\mathpalette\@versim>}$}\fi}
\def\lsim{\ifmmode{\mathrel{\mathpalette\@versim<}}
    \else{$\mathrel{\mathpalette\@versim<}$}\fi}
\def\@versim#1#2{\lower 2.9truept \vbox{\baselineskip 0pt \lineskip
    0.5truept \ialign{$\m@th#1\hfil##\hfil$\crcr#2\crcr\sim\crcr}}}
\catcode`\@=12
\documentstyle{l-aa}

\begin{document}


   \thesaurus{(07.06.1;07.14.1;07.23.1)}

   \title{Stellar systems following the $\dvm$ luminosity law.}

   \subtitle{II. Anisotropy, Velocity Profiles, and the FP of 
                 elliptical galaxies}

   \author{L. Ciotti\inst{1} and B. Lanzoni\inst{2}}

   \offprints{L. Ciotti; e-mail ciotti@astbo3.bo.astro.it}

   \institute{\inst{1}Osservatorio Astronomico di Bologna,
              via Zamboni 33, 40126 Bologna, Italy\\
              \inst{2}Dipartimento di Astronomia, Universit\`a di Bologna, 
               Via Zamboni 33, 40126 Bologna, Italy}

   \date{Received ... ; accepted ...}

   \maketitle

   \begin{abstract}

Following a first paper on this subject (Ciotti 1991, hereafter Paper I), we 
study the dynamical properties of spherical galaxies with surface luminosity 
profile described by the $\dvm$-law, in which a variable degree of orbital 
anisotropy is allowed. The parameter $m$ for the present models covers the 
range $[1,10]$. For these models we study the self-consistently generated 
phase-space distribution function (DF), and we derive -- as a function of $m$ 
-- the minimum value of the anisotropy radius for the model consistency (i.e., 
in order to have a nowhere negative DF). Then we study the region in the 
parameter space where the $\dvm$ models are likely to be stable against 
radial-orbit instability, and we compare its size with that of the larger 
region corresponding to the consistency requirement. For stable anisotropic 
models the spatial and projected velocity dispersion profiles are obtained 
solving the Jeans equation, and compared to those of the globally isotropic 
case, already discussed in Paper I. The relevance of the results in 
connection with the Fundamental Plane (FP) of elliptical galaxies is pointed 
out: the effect on the projected velocity dispersion due to the maximum 
orbital anisotropy allowed by the stability requirement is well within the
FP thickness, and so no fine-tuning for anisotropy is required. 
Finally, the Velocity Profiles are constructed as function of the projected 
radius and for various degrees of anisotropy, and their deviations from a 
gaussian discussed.


   \end{abstract}

%

\section{Introduction}
  
The $R^{1/4}$-law [Eq. (1) below, with $m=4$] was introduced by de Vaucouleurs
(1948) to describe the projected luminosity density (or surface brightness) 
$I(R)$ of elliptical galaxies, and has worked remarkably well. It has no 
{\it free parameters} and depends on two well defined {\it physical scales}: a
characteristic linear scale, $\re$, and a surface brightness factor, $\Iz$. 

A natural generalization of this empirical law was first proposed by Sersic 
(1968), as the $\dvm$-law. From an observational point of view, the $\dvm$-law 
has been widely used (see, e.g., Davies et al. 1988; Capaccioli 1989; Makino et
al. 1990; Young \& Currie 1994; Andredakis, Peletier \& Balcells 1995; 
Courteau, de Jong \& Broeils 1996). 
In particular, for normal ellipticals and brightest cluster galaxies, a 
correlation between their luminosity $L$ and the value of $m$ has been found, 
in the direction of $m$ increasing with increasing $L$ 
(Caon, Capaccioli \& D'Onofrio 1993; Graham et al. 1996). 

From a theoretical point of view much less work has been done on the 
$\dvm$-law, its apparent universality, and its possible applications to the
problem of the FP of elliptical galaxies (Djorgovski \& Davis 1987; Dressler 
et al. 1987; Bender, Burstein \& Faber 1992) and only {\it one-component, 
spherically symmetric, globally isotropic models} have been studied (Paper I; 
Hjorth \& Madsen 1991; Ciotti, Lanzoni \& Renzini 1996; Graham \& Colless 
1996). Considering the extensive use of the $\dvm$-law we plan to extend the
investigation of this class of models. 

The interest in the study of the dynamical properties of the $\dvm$ models is 
also renewed following recent ground based observations (M{\o}ller, Stiavelli,
\& Zeilinger 1995), and Hubble Space Telescope ones, showing that the spatial 
luminosity distributions of elliptical galaxies approach the power-law form 
$\rho(r)\propto r^{-\gamma}$ at small radii, with $0\leq\gamma\leq 2.5$ (Crane
et al. 1993; Jaffe et al. 1994; Ferrarese et al. 1994; Lauer et al. 1995; 
Kormendy et al. 1995; Byun et al. 1996; de Zeeuw \& Carollo 1996). 
Since the deprojected density of the 
$\dvm$ models increases toward the center as $r^{-(m-1)/m}$ for $m>1$ 
(Paper I), this family of models can be used to study power-law galaxies with 
$0<\gamma <1$.

In particular, in this paper we study one-component, spherically symmetric,
anisotropic $\dvm$ models, in which orbital anisotropy follows the widely used
OM parameterization (Osipkov 1979, Merritt 1985). We numerically construct 
their phase-space DF, and we determine the region in the parameter space where
anisotropic $\dvm$ models are consistent, i.e., their DF is positive over all 
the accessible phase-space. We then investigate the models stability against 
radial orbit instability, by using the global stability parameter, comparing 
the radial and tangential kinetic energies. In this way we approximately bound
the region in the parameter space where the anisotropic models are consistent 
but (with high probability) unstable. 

Having determined the region in the parameter space where the $\dvm$ models are
physical, we study their main properties. First of all the spatial and 
projected velocity dispersions are derived and discussed. Then we construct 
their Velocity Profiles (hereafter VPs), and fit them using the Gauss-Hermite 
series, discussing their deviations from a pure gaussian for different $m$ and
anisotropies. 

In a third paper (Paper III, Ciotti \& Lanzoni 1996) we study in detail the 
properties of the DF, VPs, and velocity dispersion profiles of two-component 
spherically symmetric $\dvm$ models, in order to understand how the 
superposition of a dark matter halo modifies the internal dynamics of the 
models and their observational properties.

The paper is organized as follows. In Sect. 2 the basic properties of the 
$\dvm$ models, already discussed in Paper I, are summarized. In Sect. 3 the DFs
for anisotropic models are derived and discussed, together with the limits 
imposed on orbital anisotropy for the models consistency and stability. In 
Sect. 4 we obtain the velocity dispersion profiles of some characteristic 
models, and some observationally interesting properties are presented. A short
discussion on the implications of the obtained results on the problem of the
FP of ellipticals is given. In Sect. 5 the models
VPs are constructed, and then fitted using the Gauss-Hermite series, thus 
quantifying the departures from gaussianity. Finally in Sect. 6 the main 
conclusions are summarized.   

\section{A summary of the properties of spherical $\dvm$ models}

Here we give a short summary of the basic properties of the $\dvm$ models, as
derived in Paper I. A full treatment is given there, and we will refer to 
equations in that paper as Eq. $(P\#)$.

The spherical $\dvm$ models are defined as a one-parameter family of stationary
stellar systems, with surface brightness profile given by
\begin{equation}
I(R)=\Iz\exp (-b\eta^{1/m}),
\end{equation}
where $\Iz$ is the central surface brightness, $\eta\equiv R/\re$, $R$ is the 
projected radius, and $\re$ is the {\it effective radius} (i.e. the projected 
radius inside which the projected luminosity equals half of the total 
luminosity). The defining parameter is $m$, a positive real number, and $b$ is
a dimensionless parameter whose value is determined by the definition of $\re$.
The function $b=b(m)$ [Eq. (P5)] is very well fitted by the linear 
interpolation $b(m)=2m-0.324$, for $0.5\leq m\leq 10$, with relative errors 
smaller than 0.001, and $b(4)=7.66924944$ (Paper I). The total luminosity 
$\ltot$ is given by $\ltot =\Iz\re ^2\times \ltota$, with
\begin{equation}
\ltota={2\pi m\over b^{2m}}\Gamma(2m),
\end{equation}
where $\Gamma$ is the complete gamma function [Eq. (P4), Erd\'ely, Magnus, 
Oberhettinger \& Tricomi 1953, hereafter EMOT, vol. I, p. 1]. 

The most important deprojected quantity associated to $I(R)$ is the 
{\it luminosity density} $\nu$, which is related to the mass density via 
$\rho (r)=\ml\nu (r)$, where $\ml$ is the mass-to-light ratio, and $r$ is the 
spatial radius. We assume a constant $\ml$, so that the main quantities (mass 
inside $r$, potential, velocity dispersion, etc.) depend only on the luminosity
density $\nu (r)$, which is related to the surface brightness profile by an 
Abel integral equation (see, e.g., Binney \& Tremaine 1987, hereafter BT):
\begin{equation}
\nu (r)=-{1\over \pi}\left [
\int_{r}^{\infty}{dI\over dR}{dR\over\sqrt{R^2 - r^2}}-\lim_{R\to\infty}
{I(R)\over\sqrt{R^2-r^2}}
\right].
\end{equation}
The second term in the r.h.s. of Eq. (3) is zero for any positive value of $m$,
and the resulting luminosity density $\nu _m(r)=({\Iz/\re})\times\dm (s)$, 
where $s\equiv r/\re$, is extensively discussed in Paper I. The asymptotic 
behaviour of $\dm$ for $r\to\infty$ is given in Eq. (P8), while for $r\to 0$ 
one obtains: 
\begin{equation}
\dm(0)={b^m\over\pi}\Gamma(1-m),\quad\quad\quad m<1
\end{equation}
\begin{equation}
\dm(s)\sim {b\over\pi}\ln\left({2\over bs}\right), \quad\quad\quad m=1
\end{equation}
\begin{equation}
\dm(s)\sim {B[1/2,(m-1)/2m]\over 2mb^{m-1}}\exp(-bs^{1/m})
s^{(1-m)/m},
\end{equation}
for $m>1$, and where $B(x,y)$ is the complete beta function 
[Eqs. (P9)-(PA5), EMOT, vol. I, p. 9]. It 
should be noted that for $m>1$ the density diverges at the origin as 
$r^{(1-m)/m}$; therefore the divergence is worse for higher-$m$ models. 
Finally, as in Paper I, we consider the {\it relative} potential $\pot (r)=
G\ml\Iz\re\times\pota (s)$, where $G$ is the gravitational constant. 
Unfortunately $\pota (s)$ cannot be expressed in terms of elementary functions,
but, at variance with the density, for $r=0$ it converges for all $m$: 
\begin{equation}
\pota (0)={4\Gamma (1+m)\over b^m},
\end{equation}
[Eq. (P12)].

\section{The DF for anisotropic $\dvm$ models}

For any collisionless stationary system the DF $f$ depends on the phase-space 
coordinates only through the isolating integrals of motion admitted by the 
underlying potential ({\it Jeans Theorem}, Chandrasekhar 1942), and moreover, 
if the system is also spherically symmetric in all its properties, $f$ depends 
only on the binding energy and on the angular momentum square modulus $L^2$.
Usually, the negative value of the binding energy, the {\it relative binding 
energy} $\e$, is used. For spherical models with $f=f(\e,L^2)$, the tangential
components of the velocity dispersion tensor are identical, the only possible 
difference being between $\srad^2$ and $\sigma _{\theta }^2=\sigma _{\phi }^2=
\stan^2/2$, and the total velocity dispersion is $\sigma^2(r)=\srad^2(r)+
\stan^2(r)$. 

In the OM formulation the radially anisotropic case is obtained assuming a DF
depending on $\e$ and $L^2$ only through the variable $Q$ defined as:
\begin{equation}
Q=\e-{L^2\over 2\ra^2},
\end{equation}
where $\ra$ is the so-called {\it anisotropy radius}, and where $f(Q)\equiv 0$ 
for $Q<0$. Under this assumption the models are characterized by radial 
anisotropy increasing with the galactic radius, and 
\begin{equation}
\beta(r)\equiv 1-{\stan^2(r)\over 2\srad^2(r)}={r^2\over r^2 + \ra^2}.
\end{equation}
In the limit $\ra\to\infty $ the velocity dispersion tensor is globally 
isotropic. The simple relation between energy and angular momentum prescribed 
by Eq. (8) allows to express the DF as:
\begin{equation}
f(Q)={1\over\sqrt{8}\pi^2}{d\over dQ}\int_0^Q {d\varrho\over d\psi}
{d\psi\over\sqrt{Q-\psi}},
\end{equation}
where 
\begin{equation}
\varrho(r)\equiv \left(1+{r^2\over\ra^2}\right) \rho(r),
\end{equation}
(BT, p.240). For ease of comparison we will use in the following $\sa\equiv
\ra/\re$, and $\dfm (Q)=[G^3\ml\Iz\re^5]^{-1/2}\times\dfmt(\Qt)$, where 
$0\leq\Qt\equiv Q/\pot(0)\leq 1$. 

\subsection{Consistency}

The basic requirement for any physically admissible DF is its non-negativity 
over the phase-space accessible to the system, and we call {\it consistent} 
any model with a nowhere negative $f$. The DF obtained by Eq. (10) is not 
automatically consistent, because the integral inversion does not guarantee its
positivity, and so for any model one has to check the consistency: if for some
positive value of $Q$ it results $f<0$, the adopted anisotropy radius is 
inconsistent with the assumed density profile. For example, in Paper I it was 
shown that globally isotropic $\dvm$ models are consistent for all the explored
values of $m$. It is therefore of interest to investigate here the consistency
of the same family of models for various degrees of anisotropy. In Fig. 1 we 
plot $\dfmt$ vs. $\Qt$ in the case of global isotropy and strong anisotropy. A
common characteristic is that the central divergence of the DFs, present in all
globally isotropic models, is unaffected by OM anisotropy, a behaviour similar
to that analytically discussed in Ciotti (1996), and for which the same 
qualitative explanation holds. This divergence -- as for any density profile 
with total finite mass -- is not a problem: although the central phase-space 
density of the models diverges, the corresponding mass does not. 
   \begin{figure}[htbp]
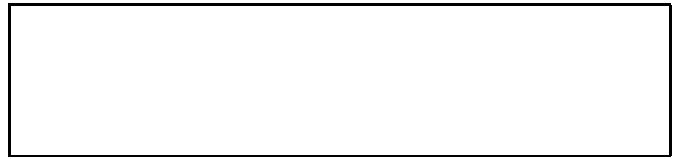

      \picplace{2cm}
      \caption{The DFs for globally isotropic (solid lines) and stable 
               anisotropic ($\xi=1.7$, dotted lines) $\dvm$ models.}
         \label{Fig1}
   \end{figure}
The requirement of consistency leads to define the critical anisotropy radius 
for consistency $\sac$ so that for $\sa <\sac$ a {\it negative} DF for some 
admissible value of $Q$ is obtained. The existence of a critical anisotropy 
radius for the $\dvm$ models is easily understandable, remembering that a {\it
completely} radial orbital distribution cannot be sustained by density profiles
less divergent than $1/r^2$ (see, e.g., Ciotti \& Pellegrini 1992 for an easy 
proof), and that the logarithmic slope for $\dvm$ models is $1/m -1$ for 
$r\to 0$. It is then interesting to show in detail the effect of a decreasing 
$\sa$ on the DF of $\dvm$ models. In particular, in Fig. 2 we show the 
modifications on the de Vaucouleurs DF. Note that approaching $\sac$ the DF 
becomes more and more depressed in the regions corresponding to intermediate 
values of the parameter $\Qt$, i.e., $\dfqt$ becomes first negative outside the
center. The dramatic effect of the anisotropy when $\sa$ is near its critical 
value is apparent in Fig. 2, where we have plotted $\dfqt$ also for a slightly
higher value of $\sa$. This behaviour is common to the whole family of the 
$\dvm$ models, and seems to be more a consequence of the OM parameterization 
itself than a characteristic of some specific mass model.\footnote{An identical
behaviour it is found also for one and two component Hernquist models, 
discussed in Ciotti (1996).}
   \begin{figure}[htbp]
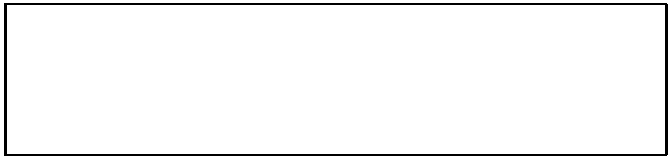

      \picplace{2cm}
      \caption{The modifications of the de Vaucouleurs DF, 
               moving from the globally isotropic case (solid line), to the 
               $\xi=1.7$ anisotropy (dotted line), and to the critical 
               anisotropy for consistency (dashed line).}
         \label{Fig2}
   \end{figure}

In Fig. 3 (solid line) $\sac$ is plotted in the parameter space $(m;\sa)$. Note
how $\sac$ asymptotically decreases towards very small values for increasing 
$m$. The qualitative trend of $\sac$ is due to the behaviour of $\dm$: for 
small $m$ this results in a quite flat density distribution, and so only "high"
values for $\sa$ are permitted; the opposite is true for high $m$ models, for 
the density becomes more and more similar to a profile $\propto 1/r$, and a 
stronger radial anisotropy is admitted. The flattening of the curve $\sac$ for
high $m$ is explained by the same argument.
   \begin{figure}[htbp]
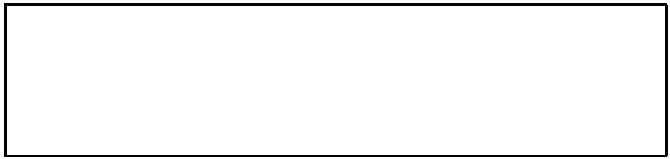

      \picplace{2cm}
      \caption{The minimum value of the anisotropy radius for the model
               consistency (solid line), and that for the model stability, for
               two different values of $\xi$. The lines are the interpolating 
               functions given in the text.}
         \label{Fig3}
   \end{figure}
As can be seen in Fig. 3, a good fit of $\sac$ as function of $m$ is given by:
\begin{equation}
\sac\simeq {\rm e}^{-0.93\,m}(1.32-9.85\,10^{-3}m^2+3.28\,10^{-3}m^4),
\end{equation}
while the exact values are given for integer $m$ in Table 1.
   \begin{table*}
      \caption{Critical values of $\sa$ for consistency and stability in the 
               cases $\xi=1.7$ and $\xi=2$, for integer values of $m$}.
         \label{Tab1}
      \picplace{4cm}
   \end{table*}

\subsection{Stability}

A given density model is not useful for applications on data-modelling if 
unstable. In Paper I it was shown in a rigorous way that globally isotropic 
$\dvm$ models are stable. Unfortunately for anisotropic models the same 
approach is not possible, and so only approximate results can be obtained 
(unless one performs a much more complex linear stability analysis). Here, 
as in Carollo, de Zeeuw, \& van der Marel 
(1995, hereafter CZM), the stability of anisotropic models is investigated in 
a semi-quantitative way using the radial-orbit instability indicator $\xi\equiv
2\Krad/\Ktan$ (see, e.g., Fridman \& Polyachenko, 1984), where $\Ktan=2\pi\int
\rho\stan^2 r^2dr$ and $\Krad=2\pi\int\rho\srad^2 r^2dr$ are the tangential and
the radial kinetic energies, and have been numerically computed. This parameter
is known to be a robust indicator, i.e., it is quite independent of the assumed
density distribution profile, and when $\xi\gsim 1.5\div 2$ the model is 
likely to be unstable. Note that for any globally isotropic model $\xi=1$, 
because $2\Krad=\Ktan$, while in presence of radial anisotropy $2\Krad >\Ktan$,
and so $\xi>1$. 
For the investigated models $\xi=\xi(\sa,m)$, and for a fixed $m$ it decreases
towards unity for increasing $\sa$ (see Fig. 4), according to the previous 
discussion. So, assuming a fiducial critical value of $\xi$ for
stability, a minimum value for the anisotropy radius $\sax$ is 
obtained, i.e., all models with $\sa <\sax$ are unstable (see also point 4 in
Sect. 6).

In Fig. 3 $\sax$ is shown for two 
different $\xi$, and the corresponding values are given in Tab. 1. As it is 
intuitive, for each $m$, $\sax >\sac$: all models in the strip $\sac\leq\sa\le
\sax$ are consistent but unstable. As in the consistency analysis, also for
stability an increase of $m$ corresponds to a decrease in $\sax$: this is due 
to the fact that inside $\sa$ the orbital distribution is nearly isotropic, and
with increasing $m$ a higher fraction of the mass is contained in the central 
regions of the model, thus exerting a more efficient stabilizing influence on 
the system (Polyachenko 1987).
   \begin{figure}[htbp]
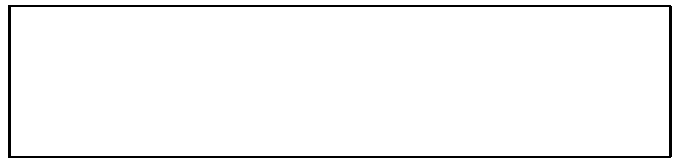

      \picplace{2cm}
      \caption{The value of the stability parameter $\xi$ for various $m$ and 
               for an increasing anisotropy radius.}
         \label{Fig4}
   \end{figure}
A fit of the minimum anisotropy radius for stability, as a function of $m$, 
and having assumed as stability requirement that $\xi=1.7$, is 
\begin{equation}
\sax\simeq {\rm e}^{-0.019m^2}(1.31-3.17\,10^{-4}m^2+1.06\,10^{-4}m^4).
\end{equation}

\section{Velocity Dispersion Profiles}

In this section we present the spatial and line--of--sight velocity dispersions
profiles for the $\dvm$ models with radial orbital anisotropy, and we compare 
them with the analogous globally isotropic cases, fully described in Paper I. 
In order to obtain the radial component $\srad (r)$, we integrate the Jeans 
equation 
\begin{equation} 
{1\over \rho (r)}{d\rho (r)\srad^2 (r)\over dr}+2{\beta(r)\srad^2 (r)\over r}
={d\psi (r)\over dr},
\end{equation}
with the natural boundary condition $\rho\srad^2\to 0$ for $r\to\infty$. 
Having assumed OM anisotropy, the integral solution can be written 
explicitly, as 
shown by Binney \& Mamon (1982), and after normalization $\srad^2(r)=
G\ml\Iz\re\times\widetilde\srad^2(s)$. The tangential velocity dispersion is
then obtained from Eq. (9). In Fig. 5, $\srad/\svir$ and 
$\sigma_{\theta}/\svir$ are shown
for some values of $m$, where $\svir^2=(2\pi/M)\int_0^{\infty}\rho\psi r^2dr$,
is the virial velocity dispersion. After normalization it results that a very
good fit is given by
\begin{equation}
\tilde\svir ^2\simeq 4.7{\rm e}^{-1.82\,m}.
\end{equation}

   \begin{figure*}
   \picplace{2cm}
   \caption{The isotropic (solid lines), radial (dotted lines), and 
            one-dimensional tangential (dashed lines) velocity dispersion 
            profiles for various $m$ and anisotropy radii. For each model the 
            $\sa$ is assumed to be the minimum possible for stability (with 
            $\xi =1.7$), and its value is the number printed in each panel.}
         \label{Fig5}
    \end{figure*}
A common feature of all the models is the characteristic central depression of 
the velocity dispersion profiles: the explanation of this behaviour for 
isotropic models was qualitatively given by Binney (1980) for the de 
Vaucouleurs law, and analytically for all the $\dvm$ models in Paper I. We note
that with increasing $m$ the maximum moves towards smaller and smaller radii 
(for the $m=10$ model the maximum is inside $s=10^{-2}$), and its value becomes
correspondingly higher and higher. The same behaviour is shown also in the 
strongly {\it anisotropic} models, both in $\srad$ and $\stan$. 
It can be noted how the position and the value of the maximum are not strongly
affected by anisotropy.

\subsection{Projected and Aperture Velocity Dispersions}

More important for observational purposes is the line--of--sight (or projected)
velocity dispersion profile $\sp (R)$, obtained from $\srad$ as: 
\begin{equation}
I(R)\sp^2 (R)=2\int _R ^{\infty} 
\left [1-\beta(r){R^2\over r^2}\right ]
{\nu (r)\srad^2 (r)rdr\over \sqrt {r^2 - R^2}},
\end{equation}
(see, e.g., BT, p. 208).

In Fig. 6 $\sp/\svir$ corresponding to the same models described in Fig. 5 is 
shown. Note that, as a consequence of projection, the central depression is 
somewhat reduced but does not completely disappear. 
   \begin{figure*}
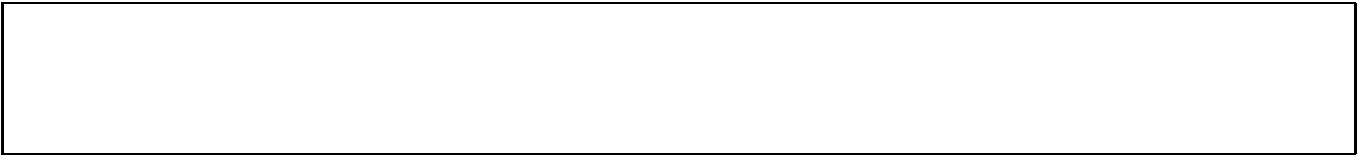

   \picplace{2cm}
   \caption{The projected velocity dispersion for the same models shown in
            Fig. 5. The solid line refers to the globally isotropic models, and
            the dotted line to the anisotropic ones.}
         \label{Fig6}
    \end{figure*}
We give here a simple approximation of the position of the maximum and its
height, as a function of $m$, in the case of global isotropy:
\begin{equation}
\left({\sp\over\svir}\right)_{\rm Max}\simeq 0.62\,{\rm e}^{0.07\,m},
\end{equation}
\begin{equation}
{R_{\rm Max}\over\re}\simeq 
{\rm e}^{-0.13m^2}(0.36-2.27\,10^{-3}m^2+7.58\,10^{-4}m^4).
\end{equation}
These fitting formulae may be useful when correcting the observed values of the
velocity dispersion, in simple applications of the virial theorem. 

When observed through an aperture of finite size, the projected velocity 
dispersion profile is weighted on
the brightness profile $I(R)$. As in Ciotti et al. (1996), we approximate this
quantity calculating the {\it aperture} velocity dispersion, defined as
\begin{equation}
LP(R)\sap^2(R)=2\pi\int_0^{R}I(R')\sp^2(R')R'\;dR',
\end{equation}
where $LP(R)$ is the projected luminosity inside $R$ [Eq. (P3)]. In Fig. 7 we
plot $\sap (R)/\svir$ for the same models of Fig. 5. Note how, independently of
the anisotropy radius, $\sap\to\svir/\sqrt{3}$ for $R\to\infty$: this result 
can be proved to be true for {\it any} assumed anisotropy (see, e.g., Ciotti 
1994). 
   \begin{figure*}
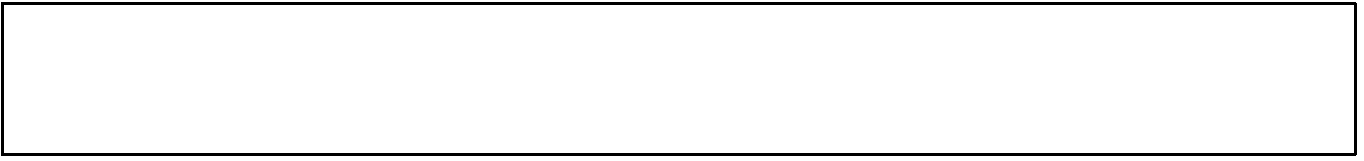

   \picplace{2cm}
   \caption{The aperture velocity dispersion for the same models shown in
            Fig. 5. The solid line refers to the globally isotropic models, and
            the dotted line to the anisotropic ones.}
         \label{Fig7}
    \end{figure*}

\subsection{Implications on the FP}

Looking at our results on the velocity dispersion profiles, we are tempted to 
discuss qualitatively their implications on the problem of the FP tilt and 
thickness.
 
Two main considerations can be made. The first concerns the effect of radial
anisotropy as a possible origin of the FP tilt, through a systematic increase 
of it with galaxy luminosity. From Fig. 7, where the maximum degree of (OM) 
anisotropy consistent with stability is considered, it is clear that the radial
anisotropy cannot produce -- in the assumption of structural homology (i.e., 
the same $m$ for all galaxies) -- the required variation of a factor of 3 in 
the observed squared velocity dispersion. This conclusion was already reached 
for different galaxy models in Ciotti et al. (1996). In any case -- at least in
principle -- a possible observational test for the importance of anisotropy 
would be to construct the FP using $\sap$ measured at large radii, and 
see whether its tilt is reduced. 

The second point concerns the problem of the very small thickness of the FP. 
From Fig. 7 one can see that for increasing $m$ the maximum deviation between 
the isotropic and anisotropic velocity dispersions becomes smaller and smaller:
for example, the percentage difference of their squared values for $m=2$ is 
$\simeq 15\%$, for $m=4$ it is already reduced to $\simeq 7\%$, and for $m=10$ 
is less than $6\%$. This implies that the anisotropy is not required to be 
fine-tuned with the galaxy luminosity in order to maintain the small observed 
FP scatter ($<12\%$ in $\sap^2$; see, Ciotti et al. 1996); on the contrary, 
all the admissible anisotropies can be present at each luminosity maintaining 
at the same time the FP thin.

\section{Velocity Profiles}

The velocity profile (VP) at a certain projected distance from the galaxy 
center is the distribution of the stars line--of--sight velocities at that 
point. It is strictly linked to the line profile in the observed spectrum, 
that is the convolution of the stellar spectra with a certain VP. The 
shape of the VPs depends not only on the galactic potential, but also on the 
stars orbital distribution, a dynamical property not fully determined by the 
galactic potential itself. That is why the usual assumption of a gaussian shape
for the VPs can generate a loss of information, and for this reason it has been
suggested that the analysis of the deviations of VPs from gaussianity may give
important insights on the dynamical structure of a galaxy (van der Marel 1994).
 
In the case of OM anisotropy the VPs can be numerically recovered from the 
$f(Q)$ using the integral expression given by CZM. As usual, we expand the VPs
on the Gauss-Hermite basis:
\begin{equation}
{\rm VP}(v)=
{\gamma \,{\rm e}^{-v^2\over{2\,\sigma^2}}\,\over \sqrt{2\,\pi}\,\sigma}
\sum_{j=0}^N h_j\,H_j(v/\sigma),
\end{equation}
(Gerhard 1993, van der Marel \& Franx 1993), where the $H_j$ are the Hermite 
polynomials as given in van der Marel \& Franx. The even coefficients $h_{2j}$
measure symmetric deviations from a pure gaussian, while the odd coefficients 
$h_{2j+1}$ are identically zero for our models because the DF depends only 
on $L^2$.
We limit our investigation to the coefficient $\hq$, because higher order 
coefficients are usually not available from spectroscopic observations. The 
uncertainties in the published data on $\hq$ are of the order of $\sim 0.02$ 
(CZM). In practice, we fix $h_0=1$, and $h_2=0$ in Eq. (20), thus requiring 
that the first term is the best-fitting gaussian, and we minimize the $\chi^2$
using the Levenberg--Marquardt method ({\it Numerical Recipes}, p.678) for the
simultaneous non linear fit of $(\gamma,\sigma,\hq)$. 
   \begin{figure*}
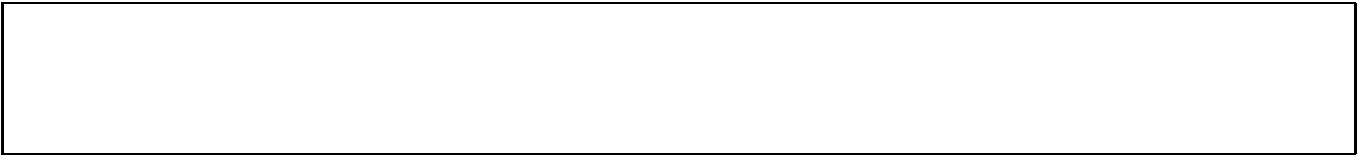

   \picplace{2cm}
   \caption{The values of the parameters $\hq$ (upper panels) and $\sigma$ 
            (normalized to $(G\ml\Iz\re)^{1/2}$, lower panels) characterizing 
            the VPs of the $\dvm$ models, as a function of $R$, for $m=1$ 
            (circles), $m=2$ (crosses), $m=4$ (squares), and $m=10$ 
            (triangles). Left panels refer to the isotropic models, right 
            panels to anisotropic ($\xi=1.7$) models.}
         \label{Fig8}
    \end{figure*}
In Fig. 8 the radial behaviour of $\sigma$ and $\hq$ for various $m$ and for 
isotropic (left panels) and anisotropic ($\xi=1.7$, right panels) models are 
shown.

Isotropic $\dvm$ models have surprisingly gaussian VPs, as the smallness of 
their $\hq$ indicates (Fig. 8a). Exception is made by the low-$m$ models, 
which present strongly non--gaussian VPs at small radii. Note also how, for 
$R\gsim 0.7\re$ the $\hq$ are completely indistinguishable for all $m$ and 
indicate (slightly) flat-topped VPs (as generally a negative $\hq$ indicates). 
Moving inside, the differences between models become more and more important,
but the global trend is towards VPs more peaked than the best-fitting gaussian,
especially for low values of $m$. The radial behaviour of the corresponding
dispersions $\sigma$ of the best-fitting gaussian is shown in Fig. 8b, where 
the similarity (apart a re-scaling) with the dispersions plotted in Fig. 6 is 
evident. 

For anisotropic models the $\sigma$ (Fig. 8d) are very similar to that of the 
isotropic case, while the situation is quite different for the $\hq$ (Fig. 8c).
Their values systematically decrease with radius in the inner regions, but
rapidly increase at large radii, indicating significantly top--peaked VPs.
This general trend, and the values as well, are nearly the same as those found
by CZM for the $\gamma$ models, and are due to the orbital distribution in the 
outer part of models with OM radial anisotropy.

Because of the differences in the trend of $\hq$ between the isotropic and the
anisotropic cases, and because of the growing evidence that $\dvm$-law 
appropriately describe the surface brightness profiles of elliptical galaxies,
we conclude that a detailed study of the VPs along the FP could be in principle
a tool to study the effect of orbital anisotropy on its tilt and thickness.

\section{Conclusions}

The results of this work are the following:
\begin{enumerate}

\item For OM anisotropic $\dvm$ systems the consistency region in the parameter
      space is explored, and the minimum anisotropy radius that can be assumed 
      for given $m$ in order to have a physical model is determined. We find 
      that models with high $m$ can sustain more anisotropy than models with 
      low $m$. A fit of the critical anisotropy radius for consistency is 
      given.

\item The maximum anisotropy tolerated by $\dvm$ models in order to be stable 
      against radial orbit instability is approximately derived, with the aid 
      of the standard stability parameter $\xi$. As expected, in this case the
      limitation on $\sa$ is stronger than that required by consistency. Again,
      high-$m$ models are more stable than low $m$ models. A fit of the minimum
      anisotropy radius permitted for stability is given.

\item The spatial, projected, and aperture velocity dispersions are derived for
      various degrees of anisotropy. Their main characteristic is the fact that
      the off-center maximum -- a feature already known and discussed for 
      globally isotropic $\dvm$ models -- is still present, and not very much 
      affected even by a strong anisotropy. 
    
\item The implications of this work for the problem of the tilt and thickness 
      of the FP of elliptical galaxies are that orbital anisotropy cannot be at
      the origin of the tilt if galaxies are described by the $\dvm$ law and 
      characterized by structural homology. At the same time, the small 
      thickness of the FP at fixed luminosity does not require any fine tuning 
      between anisotropy and luminosity, due to the stability requirement. 
      We note
      that the Saha (1991, 1992) works on stability further strengthen
      our conclusions: he found that radial orbit instability can affect also
      models with $\xi$ {\it smaller} than the values suggested by Fridman \&
      Polyachenko (1984) and here used, and so a still smaller amount of 
      anisotropy would be permitted.  

\item The VPs are studied at various distances from the center for different 
      anisotropy degrees and values of $m$. For globally isotropic models the 
      VPs are very well approximated by a gaussian, except for very small radii
      and for low $m$, where detectable deviations from a pure gaussian are 
      revealed. The lower order correction, parameterized by the coefficient 
      $\hq$, shows that outside $\re$ the VPs are flat--topped and essentially
      indistinguishable for different values of $m$. On the contrary, the VPs 
      of anisotropic models at $R\gsim\re$ are more centrally peaked than a 
      gaussian, and the values of $\hq$ increase with increasing $m$. 

\end{enumerate}

\begin{acknowledgements}
We would like to thank George Djorgovski for discussions, Roeland 
van der Marel for comments and Silvia Pellegrini for a careful reading of the 
manuscript. Also the anonymous referee is acknowledged for useful comments 
that improved the presentation of the paper. This work has
been partially supported by the Italian MURST.
\end{acknowledgements}


\begin{thebibliography}{}

   \bibitem{} Andredakis Y.C., Peletier R.F., Balcells M., 1995,
              MNRAS, 275, 874

   \bibitem{} Bender R., Burstein D., Faber S.M., 1992, 
              ApJ, 399, 380

   \bibitem{} Binney, J.J., 1980, 
              MNRAS, 190, 873

   \bibitem{} Binney J.J., Mamon G.A., 1982,
              MNRAS, 200, 361 

   \bibitem{} Binney J.J., Tremaine S., 1987, 
              Galactic Dynamics, 
              Princeton University Press, Princeton (BT)

   \bibitem{} Byun Y.I., Grillmair C., Faber S.M., Ajhar E.A., Dressler A., 
              Kormendy J., Lauer T.R., Richstone D.O., Tremaine S.D., 1996, 
              AJ, in press 

   \bibitem{} Caon N., Capaccioli M., D'Onofrio M., 1993,
              MNRAS, 265, 1013

   \bibitem{} Capaccioli M., 1989,
              in The World of Galaxies,
              eds. H.G. Corwin and L. Bottinelli, Springer-Verlag, Berlin

   \bibitem{} Carollo C.M., de Zeeuw P.T., van der Marel R.P., 1995, 
              MNRAS, 276, 1131 (CZM)

   \bibitem{} Chandrasekhar S., 1942, 
              Principles of Stellar Dynamics, 
              Chicago University Press, Chicago

   \bibitem{} Ciotti L., 1991,
              A\&A, 249, 99 (Paper I)

   \bibitem{} Ciotti L., 1994,
              Cel. Mech. \& Dyn. Astron., 60, 401

   \bibitem{} Ciotti L., 1996,
              ApJ, in press (November 1 issue)

   \bibitem{} Ciotti L., Pellegrini S., 1992,
              MNRAS, 255, 561

   \bibitem{} Ciotti L., Lanzoni B., Renzini, A., 1996,
              MNRAS, 282, 1

   \bibitem{} Ciotti L., Lanzoni B., 1996,
              in preparation (Paper III)

   \bibitem{} Courteau S., de Jong R.S., Broeils A.H., 1996,
              ApJL, 457, L1

   \bibitem{} Crane P., et al., 1993, 
              AJ, 106, 1371

   \bibitem{} Davies J.I., Phillips S., Cawson M.G.M., Disney M.J.,
              Kibblewhite E.J., 1988,
              MNRAS, 232, 239

   \bibitem{} de Vaucouleurs G., 1948,
              Ann. d'Astroph., 11, 247

   \bibitem{} de Zeeuw P.T., Carollo C.M., 1996
              IAU Symposium 171, New Light on Galaxy Evolution,
              eds. R. Bender and R.L. Davies,
              Dordrecht: Kluwer

   \bibitem{} Djorgovski S., Davis M., 1987, 
              ApJ, 313, 59

   \bibitem{} Dressler A., Lynden-Bell D., Burstein D., Davies R.L., 
              Faber S.M., Terlevich R.J., Wegner G., 1987, 
              ApJ, 313, 42

   \bibitem{} Erd\'ely A., Magnus W., Oberhettinger F., Tricomi F.G., 1953,
              Higher transcendental functions, 
              McGraw-Hill Book Company, Inc. (EMOT)

   \bibitem{} Ferrarese L., van den Bosch F.C., Ford H.C., Jaffe W., 
              O'Connell R.W., 1994, 
              AJ, 108, 1598

   \bibitem{} Fridman A.M., Polyachenko V.L., 1984, 
              Physics of Gravitating Systems, 2 vols.
              New York: Springer 

   \bibitem{} Gerhard O.E., 1993,
              MNRAS, 265, 213 

   \bibitem{} Graham A., Lauer T.R., Colless M., Postman M., 1996, 
              ApJ, 465, 534

   \bibitem{} Graham A., Colless M., 1996, preprint

   \bibitem{} Hjorth J., Madsen J., 1991,
              MNRAS, 253, 703

   \bibitem{} Jaffe W., Ford H.C., O'Connell R.W., van den Bosch F.C., 
              Ferrarese L., 1994, 
              AJ, 108, 1567

   \bibitem{} Kormendy J., Byun Y.I., Ajhar E.A., Lauer T.R., Dressler A., 
              Faber S.M., Grillmair C., Gebhart K., Richstone D.O., 
              Tremaine S.D., 1995, 
              IAU Symposium 171: New Light on Galaxy Evolution, p. 105,
              eds. R. Bender and R.L. Davies, 
              Dordrecht: Kluwer

   \bibitem{} Lauer T.R., Ajhar E.A., Byun Y.I., Dressler A., Faber S.M., 
              Grillmair C., Kormendy J., Richstone D.O., Tremaine S.D., 1995,
              AJ, 110, 2622

   \bibitem{} Makino J., Akiyama K., Sugimoto D., 1990, 
              Publ. Astron. Soc. Japan, 42, 205

   \bibitem{} Merritt D., 1985, 
              AJ, 90, 1027 

   \bibitem{} M{\o}ller P., Stiavelli M., Zeilinger W.W., 1995, 
              MNRAS, 276, 979

   \bibitem{} Osipkov L.P., 1979, 
              Pis'ma Astron.Zh., 5, 77

   \bibitem{} Polyachenko V.L., 1987, 
              IAU Symposium 127: Structure and Dynamics of Elliptical 
              Galaxies, p. 301, ed. P.T. de Zeeuw, Dordrecht, Reidel

   \bibitem{} Press W.H., Teukolsky S.A., Vetterling W.T., Flannery B.P., 1992,
              Numerical Recipes in Fortran, Second Edition, 
              Cambridge University Press

   \bibitem{} Saha P., 1991,
              MNRAS, 248, 494

   \bibitem{} Saha P., 1992,
              MNRAS, 254, 132

   \bibitem{} Sersic J.L., 1968, 
              Atlas de Galaxias Australes,
              Cordoba: Observatorio Astronomico

   \bibitem{} van der Marel R.P., Franx M., 1993,
              ApJ, 407, 525 

   \bibitem{} van der Marel R.P., Ph.D. Thesis, 1994

   \bibitem{} Young C.K., Currie M.J., 1994,
              MNRAS, 268, L11

\end{thebibliography}
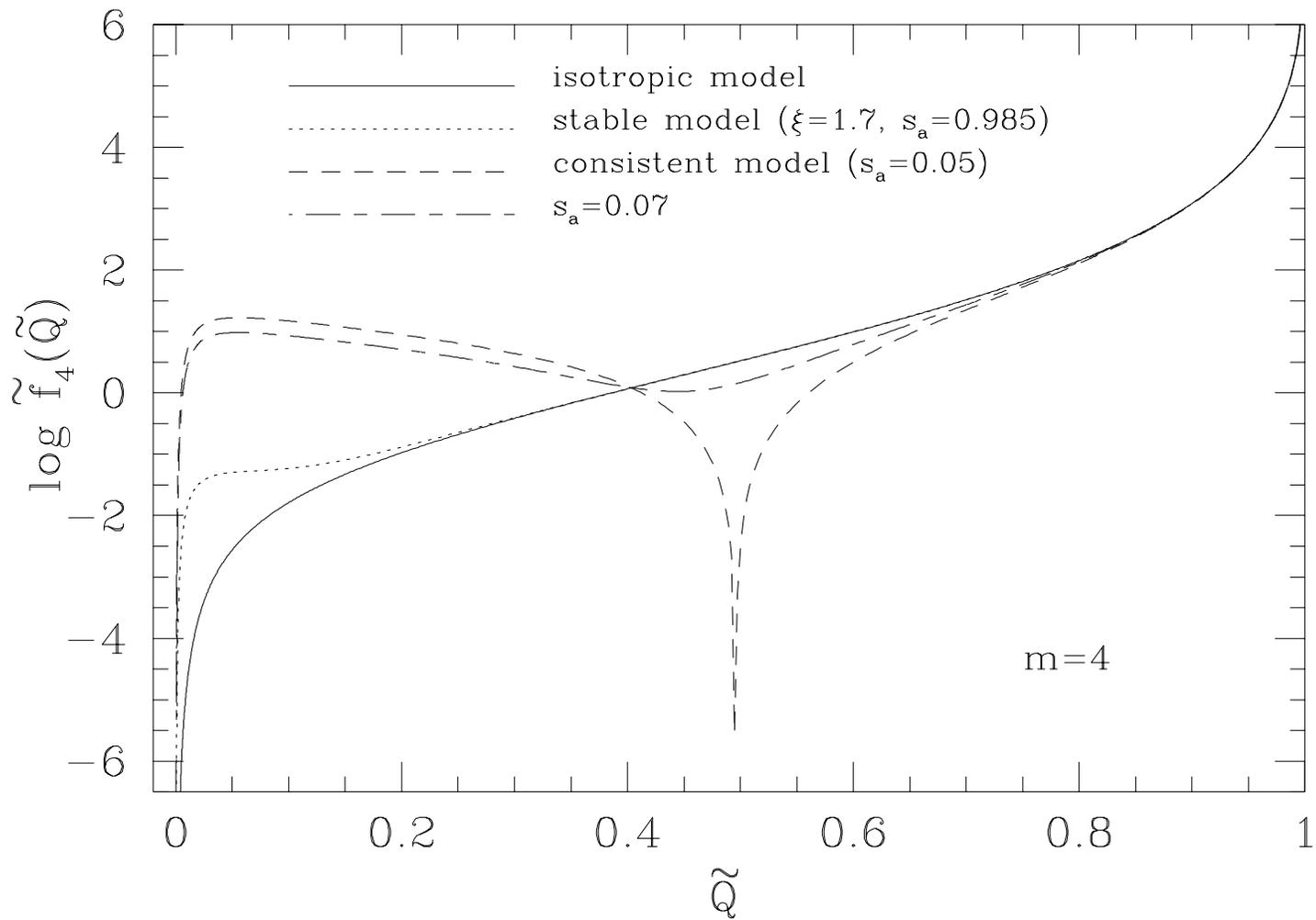
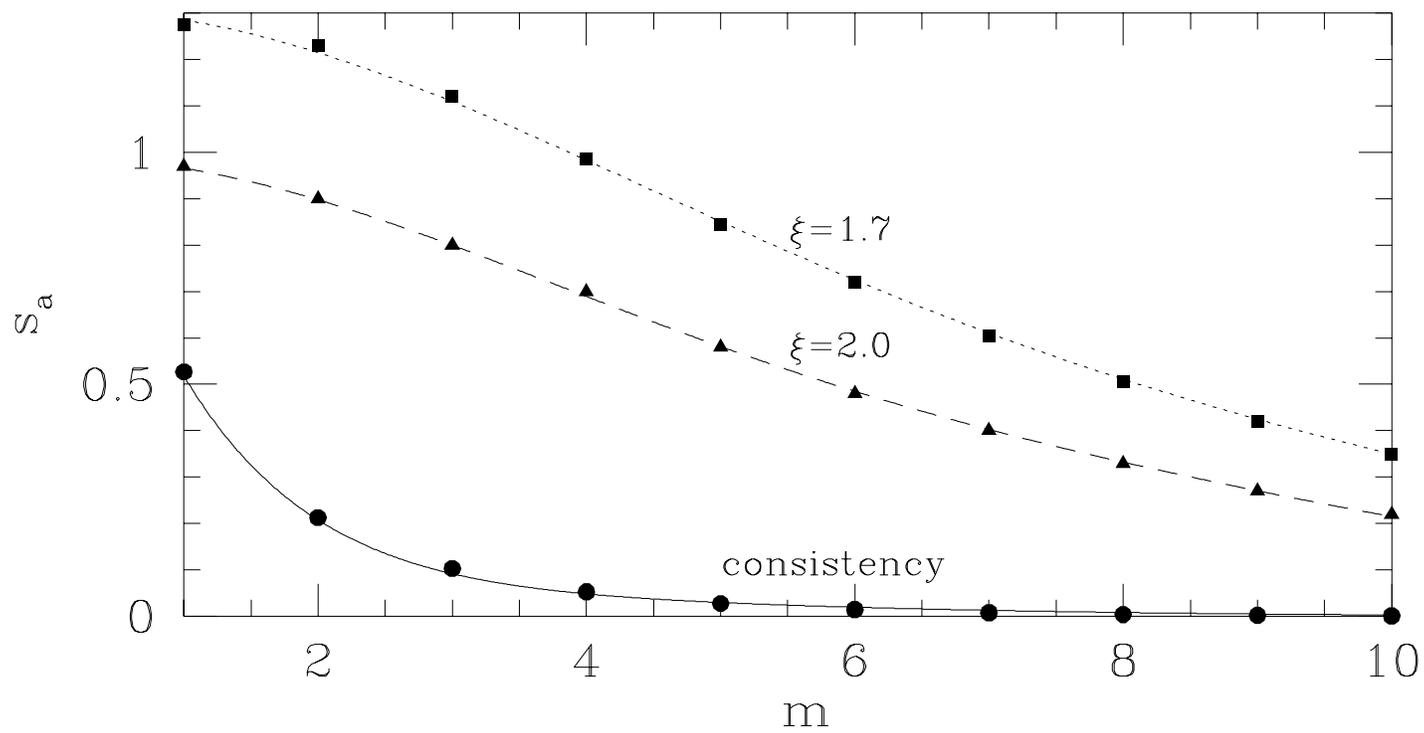
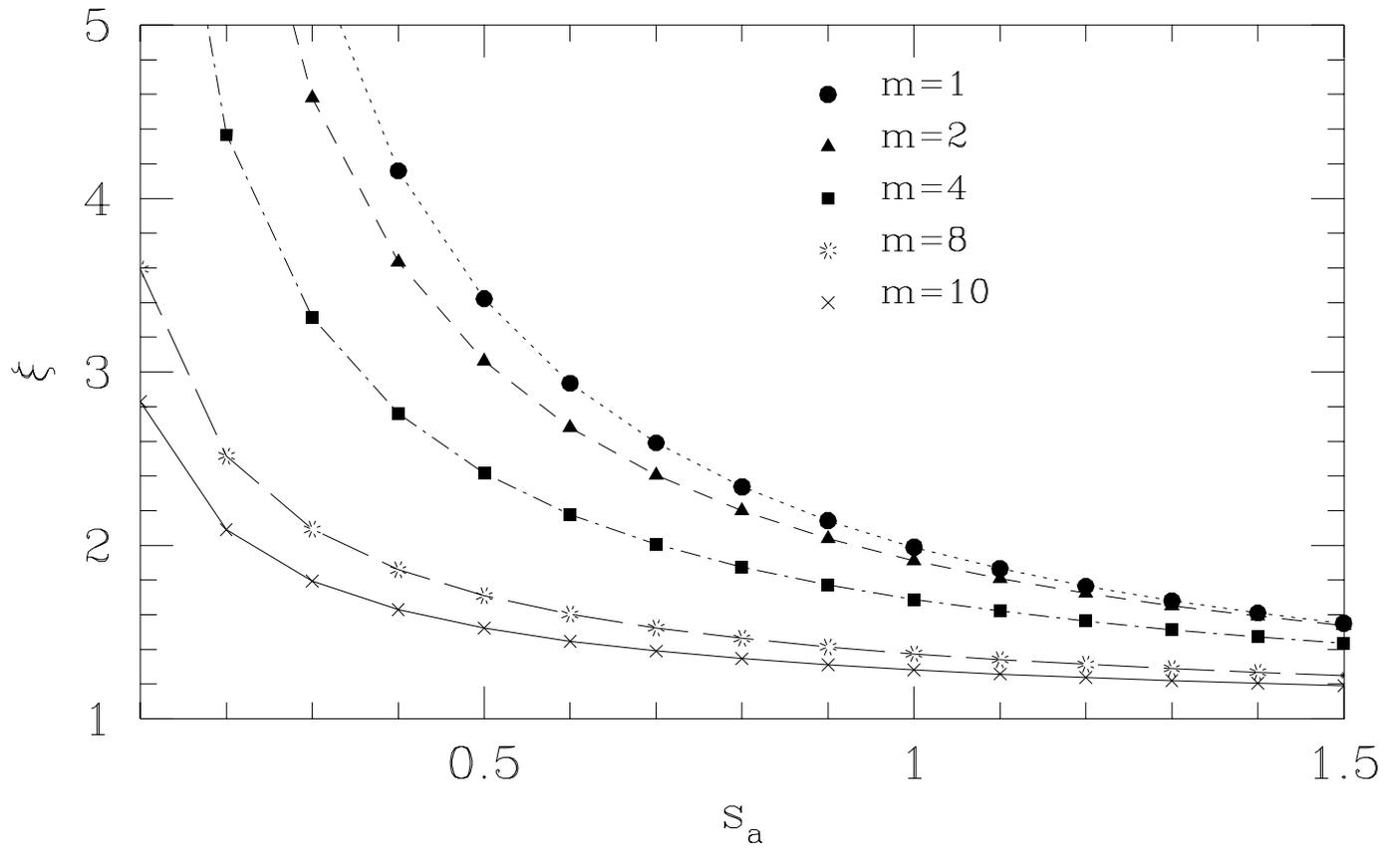
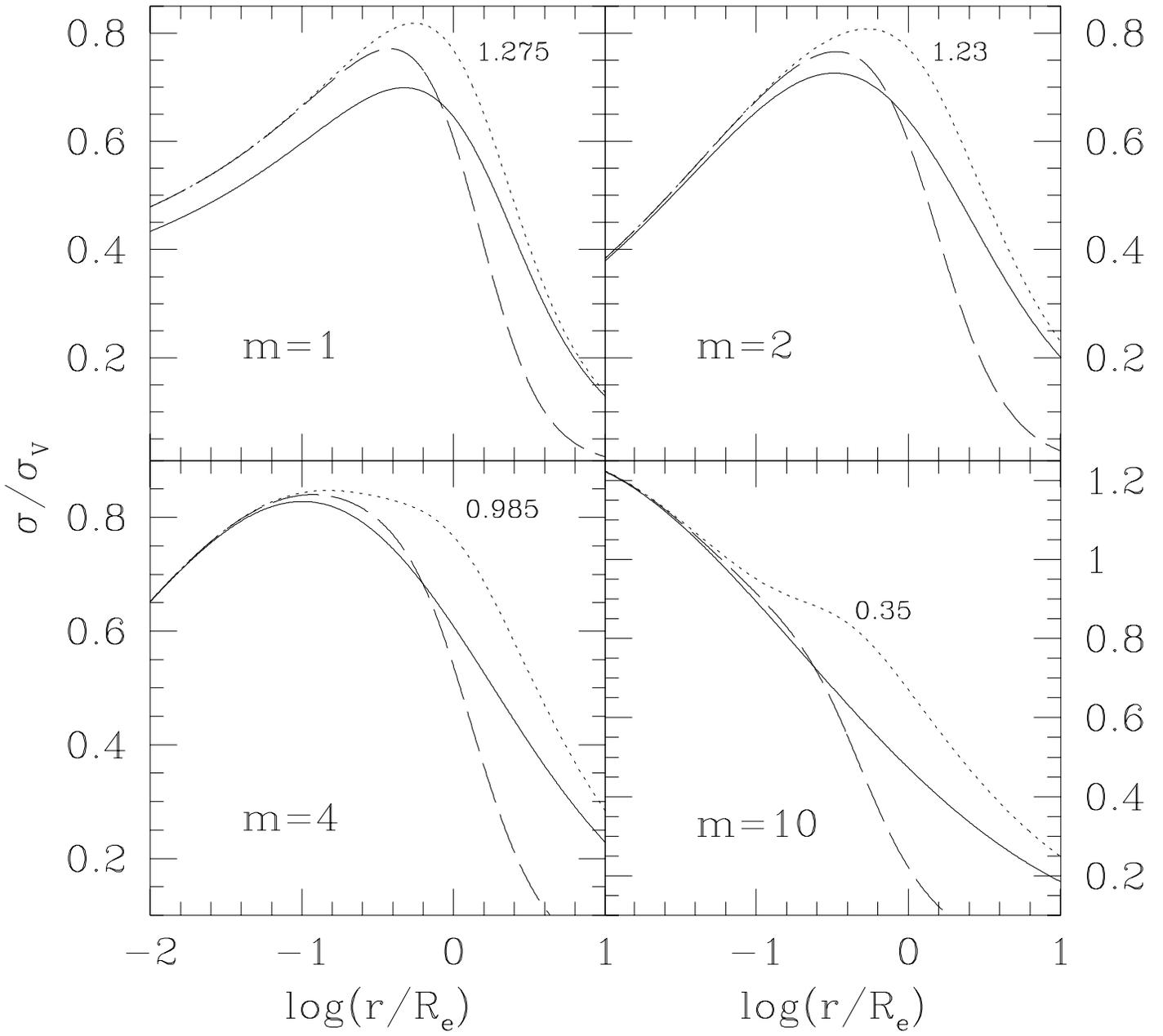
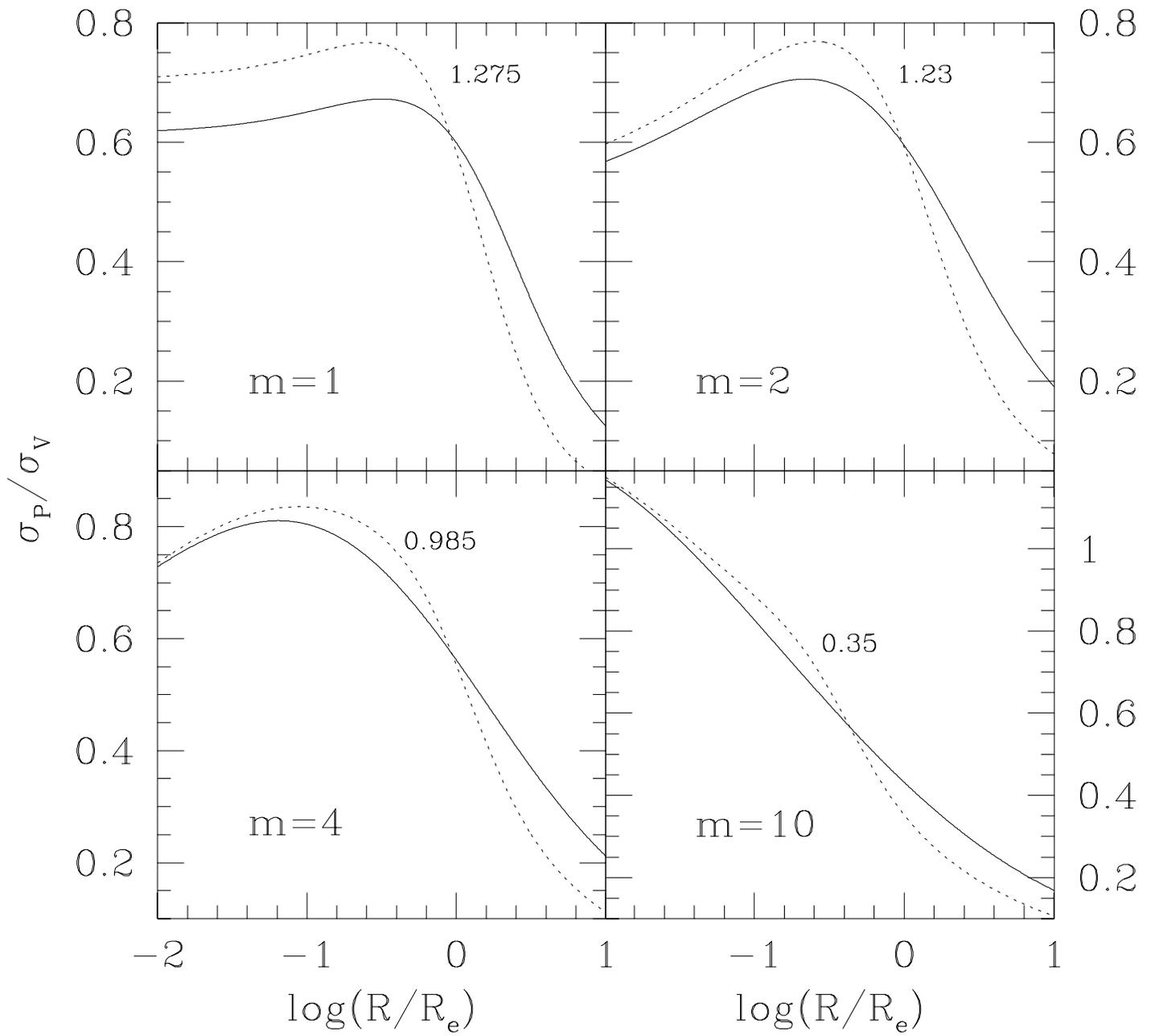
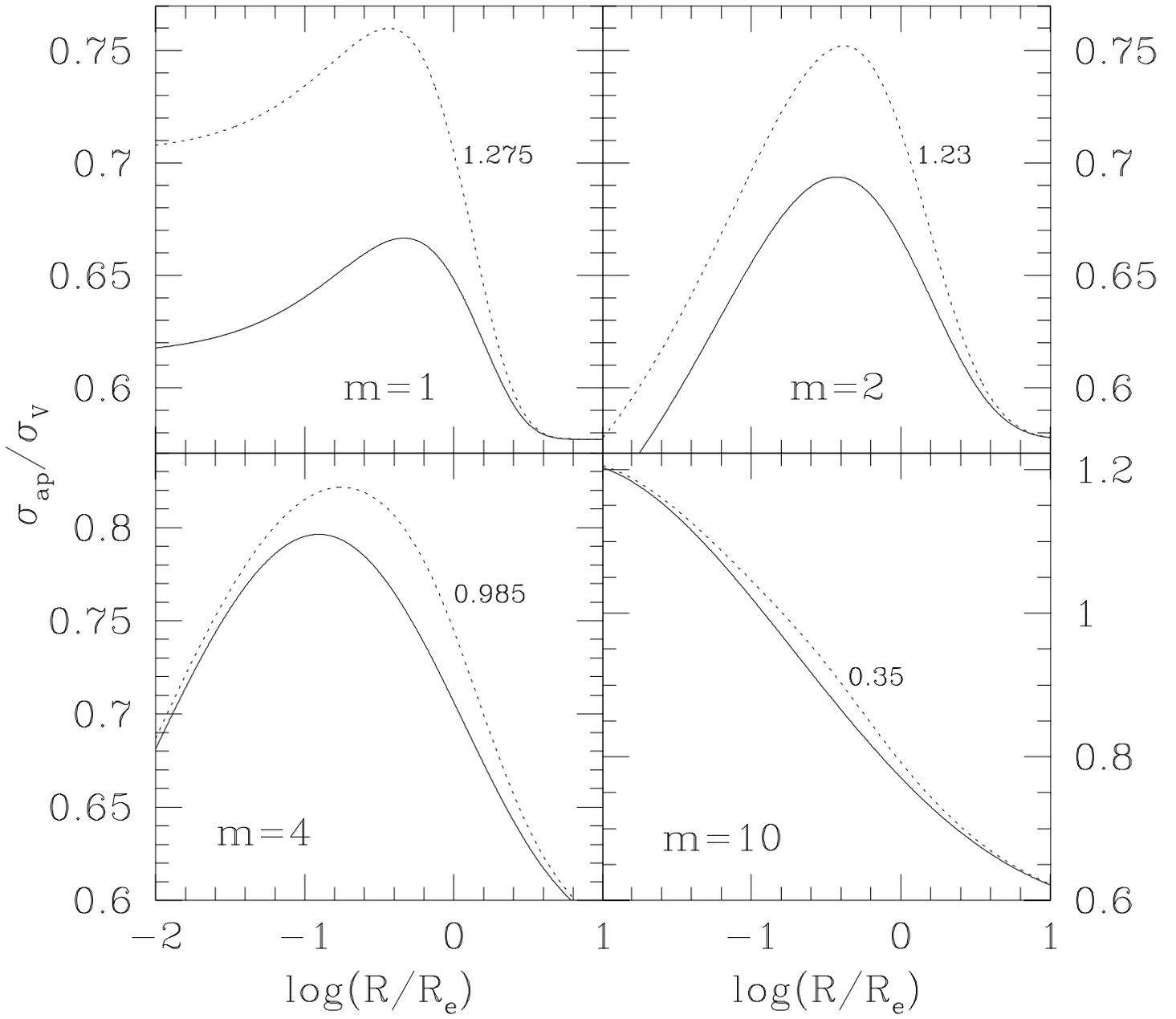
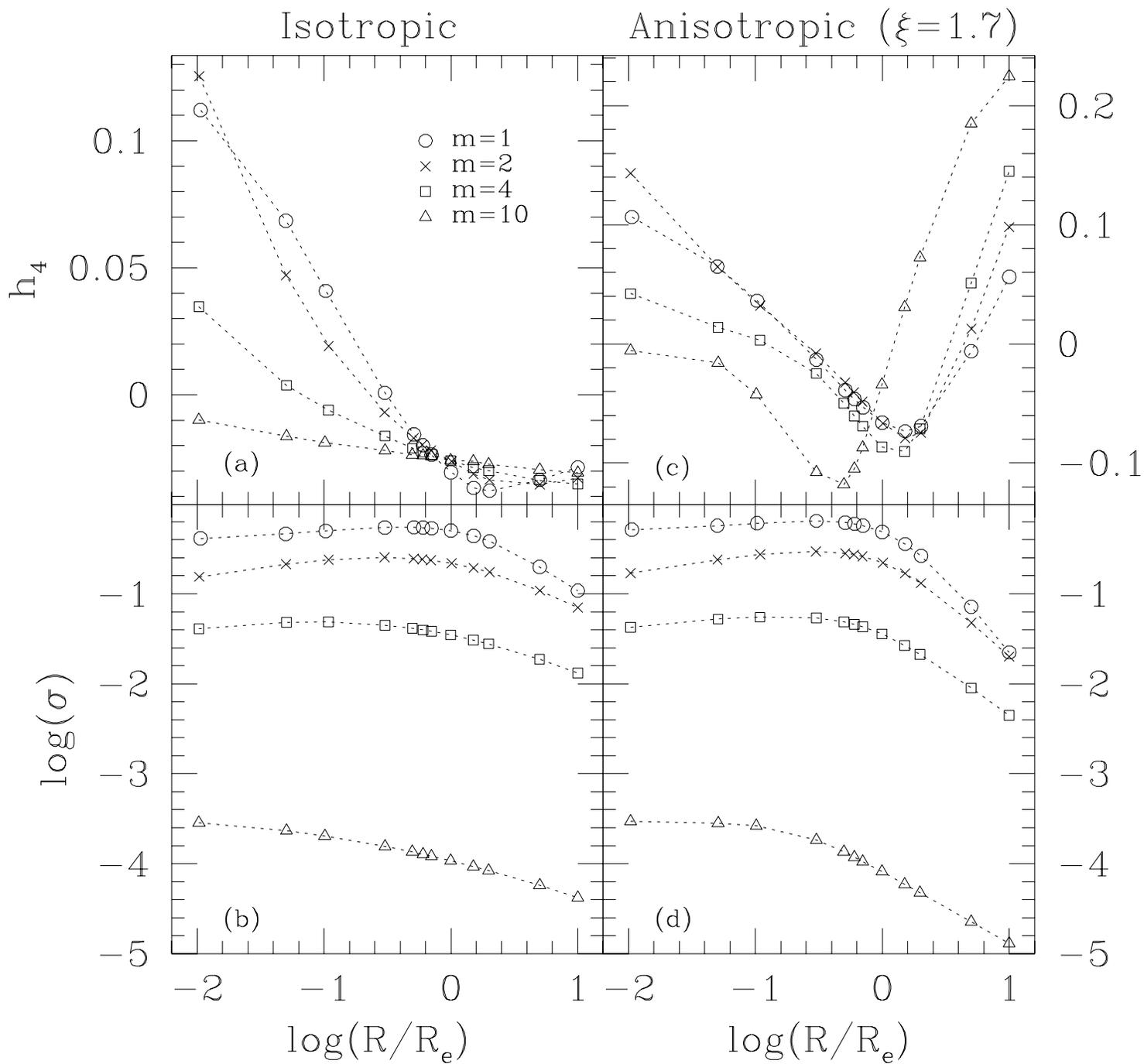
\end{document}